\documentclass[sigconf]{acmart}
\AtBeginDocument{%
  \providecommand\BibTeX{{%
    \normalfont B\kern-0.5em{\scshape i\kern-0.25em b}\kern-0.8em\TeX}}}

\copyrightyear{2023} 
\acmYear{2023} 
\setcopyright{acmlicensed}\acmConference[WWW '23 Companion]{Companion Proceedings of the ACM Web Conference 2023}{April 30-May 4, 2023}{Austin, TX, USA}
\acmBooktitle{Companion Proceedings of the ACM Web Conference 2023 (WWW '23 Companion), April 30-May 4, 2023, Austin, TX, USA}
\acmPrice{15.00}
\acmDOI{10.1145/3543873.3587571}
\acmISBN{978-1-4503-9419-2/23/04}

\usepackage{subfig}
\usepackage{multirow}
\usepackage{booktabs}

\begin{document}

\title{Retrieving false claims on Twitter during the Russia-Ukraine conflict}

\author{Valerio La Gatta}
\authornote{V. La Gatta and C. Wei contributed equally to this work.}
\orcid{1234-5678-9012}
\affiliation{%
  \institution{Information Sciences Institute, University of Southern California}
  \country{Los Angeles, California, USA}
}
\affiliation{%
  \institution{University of Naples Federico II}
  \city{Naples}
  \country{Italy}
}

\email{valerio.lagatta@unina.it}

\author{Chiyu Wei}
\authornotemark[1]
\affiliation{%
  \institution{Information Sciences Institute \\ University of Southern California}
  \city{Los Angeles}
  \state{California}
  \country{USA}}
\email{chiyuwei@isi.edu}
\author{Luca Luceri}
\affiliation{%
  \institution{Information Sciences Institute \\ University of Southern California}
  \city{Los Angeles}
  \state{California}
  \country{USA}}
\email{lluceri@isi.edu}

\author{Francesco Pierri}
\affiliation{%
  \institution{Polititecnico di Milano}
  \country{Milan, Italy}
  }
\affiliation{%
  \institution{Information Sciences Institute, University of Southern California}
  \country{Los Angeles, California, USA}
  }
\email{francesco.pierri@polimi.it}

\author{Emilio Ferrara}
\affiliation{%
  \institution{Information Sciences Institute \\ University of Southern California}
  \city{Los Angeles}
  \state{California}
  \country{USA}}
\email{ferrarae@isi.edu}

\renewcommand{\shortauthors}{La Gatta et al.}

\begin{abstract}


Nowadays, false and unverified information on social media sway individuals' perceptions during major geo-political events and threaten the quality of the whole digital information ecosystem. Since the Russian invasion of Ukraine, several fact-checking organizations have been actively involved in verifying stories related to the conflict that circulated online. In this paper, we leverage a public repository of fact-checked claims to build a methodological framework for automatically identifying false and unsubstantiated claims spreading on Twitter in February 2022. Our framework consists of two sequential models: First, the \emph{claim detection} model identifies whether tweets incorporate a (false) claim among those considered in our collection. Then, the \emph{claim retrieval} model matches the tweets with fact-checked information by ranking verified claims according to their relevance with the input tweet. Both models are based on pre-trained language models and fine-tuned to perform a text classification task and an information retrieval task, respectively. In particular, to validate the effectiveness of our methodology, we consider 83 verified false claims that spread on Twitter during the first week of the invasion, and manually annotate 5,872 tweets according to the claim(s) they report. Our experiments show that our proposed methodology outperforms standard baselines for both \emph{claim detection} and \emph{claim retrieval}. Overall, our results highlight how social media providers could effectively leverage semi-automated approaches to identify, track, and eventually moderate false information that spreads on their platforms. 

\end{abstract}





\maketitle

\section{Introduction}
On 24th February 2022, Russia started its still ongoing invasion of Ukraine, causing unprecedented backlashes for the rest of the world.\footnote{\url{https://en.wikipedia.org/wiki/2022_Russian_invasion_of_Ukraine}} Soon afterward, concerns were raised about the presence of Russian disinformation campaigns on online social media, aimed at repurposing  the invasion as a ``special operation'' against alleged Nazis in Ukraine, or attempting to blame NATO's expansion for causing the invasion \cite{hanley2022special,park2022voynaslov,pierri2022propaganda,hanley2022happenstance,pierri2022does}. Russian interference with other countries' democratic processes is anything but new, as extensively reported in the past, especially in the context of the 2016 U.S. Presidential election \cite{badawy2018analyzing,luceri2020detecting}.
During major geo-political events, manual fact-checking represents the leading strategy to debunk false information through domain experts' analyses, crowd-sourcing approaches \cite{10.1145/3511808.3557279}, and semi-automatic systems assessing news truthfulness \cite{DBLP:journals/corr/abs-2108-11896}. However, given the impossibility of manually checking every piece of information circulating online \cite{pierri2019false}, social media providers still struggle with 
keeping track of and moderating false information that spreads online \cite{pierri2023one,nogara2022disinformation}. 

Given these premises, in this work, we aim to build a methodological framework to automatically identify false and unsubstantiated claims -- verified by news agencies and fact-checking organizations (e.g., Politifact, Snopes) -- that were shared on Twitter at the dawn of the Russian invasion of Ukraine. To this end, we collect 83 false claims that were verified in the first weeks of the invasion and annotate 5,872 original tweets based on the claim(s) they discuss. Then, we deploy an automatic pipeline that comprises two models: (i) the \emph{claim detection} model identifies whether an input tweet contains a (false) claim among the 83 present in our collection; and, assuming the input tweet reports a claim, (ii) the \emph{claim retrieval} model ranks those 83 claims according to their relevance with the input tweet. For these models, we leverage modern transformer-based architectures and adopt transfer learning on the annotated dataset to optimize the performance on both tasks. 

Previous work to design automatic tools for fact-checking follows two main directions. On the one hand, research extensively focuses on the problem of assessing the truthfulness of pieces of information \cite{DBLP:journals/corr/abs-2103-07769}, 
also supporting the veracity prediction with factual evidence \cite{samarinas-etal-2021-improving,Chen_2022}. On the other hand, relatively less research is devoted to the \emph{claim detection} problem, which can be formulated as a ranking task where several input sentences are ranked by check-worthiness \cite{DBLP:journals/corr/abs-2109-12987,10.1145/3297722}, or as a binary classification task to predict whether an input sentence constitutes a claim \cite{DBLP:journals/corr/abs-2101-11891,10.1145/3412869,10.1145/2806416.2806652}. In this paper, we consider the latter setting as we aim to identify false claims and we are less interested in prioritizing tweets that require fact-checking. 

Once verified that the input tweet reports a (false) claim, we aim to retrieve the most relevant verified claim(s) in our collection. \citet{shaar-etal-2020-known} have recently proposed and defined  \emph{claim retrieval} as the task of ranking a corpus of verified documents according to their relevance to an input text. Recently, competitors at CheckThat!2021 (\cite{DBLP:journals/corr/abs-2109-12987}) have shown that fine-tuning transformer-based models lead to promising performance improvements with respect to standard information retrieval approaches (e.g., BM25 \cite{10.1561/1500000019}). Accordingly, \citet{10.1007/978-3-030-99736-6_25} obtain similar results under multilingual COVID-19 claims. Conversely, considering a political debate scenario, \citet{DBLP:journals/corr/abs-2104-07423} evaluate the impact of modeling the claim's global and local contexts on the ranking performance. Overall, the above contributions compete on benchmark datasets and assume that input sentences contain a claim. Instead, we define a more challenging scenario where \emph{claim retrieval} is combined with \emph{claim detection} in a pipeline, whose goal is to understand whether a generic tweet reports a false claim related to the specific topic of the Russian invasion of Ukraine.

\paragraph{\textbf{Contributions of this work}}
To summarize, our contributions are as follows:

\begin{itemize}
    \item We collect 83 false claims that spread on Twitter during the first weeks of the invasion and we manually annotate 5,872 original tweets to determine whether they discuss a (false) claim and, if so, which claim(s) they discuss. 
    \item We develop an automatic pipeline to detect and retrieve tweets discussing any of the 83 false claims in our collection. In particular, our models are based on modern transformer-based architectures and perform, in order, \emph{claim detection} and \emph{claim retrieval}. 
    \item We show the effectiveness of the proposed approach in retrieving the claim(s) that are referenced in the input tweet. In addition, we also observe how our models generalize to new claims that were unseen during the training process.
\end{itemize}

\section{Methodology}

In this section, we describe the data collected for the analysis and the methodology employed to annotate tweets and their corresponding claims. Then, we present our methodological framework by formally defining the \emph{claim detection} and the \emph{claim retrieval} tasks and corresponding models.
 
\subsection{Data Collection}

\begin{table*}
    \centering

    \caption{Examples of some tweet-claim pairs annotated in the dataset}
    \label{tab:examples}
    
    \scalebox{1.}{
    \begin{tabular}{cll}\toprule
    
  \textbf{No.} & \textbf{Claim} & \textbf{Tweet}  \\
 \midrule

\multirow{3}{*}{\begin{tabular}[l]{r@{}} 1 \end{tabular}} 
& Russian President Vladimir Putin threatened & Putin has warned India that don't try to interfere in  \\
& India against getting involved in the Ukraine & their matter, otherwise be ready to face the consequences \\
& crisis. & \\
 \midrule


\multirow{3}{*}{\begin{tabular}[l]{r@{}} 2 \end{tabular}} 
& The President Of Ukraine, Volodymyr Zelenskyy, & Volodymyr Zelenskyy the president of Ukraine has \\
&  Is On The Ground With His Fellow Troops  &  decided to stay behind and fight among his people \\
&   &  against the Russian army send to kyiv $[...]$ \\ \midrule

\multirow{3}{*}{\begin{tabular}[l]{r@{}} 3 \end{tabular}} 
& The Russian armed forces are not striking at the  & It is clear that the Russian army does not want to harm  \\
&  cities of Ukraine; they are not threatening the  & civilians, its strikes were directed only at military \\
&  civilian population. & targets, $[...]$ life seems almost normal in Kiev.  \\ \midrule
\multirow{3}{*}{\begin{tabular}[l]{r@{}} 4 \end{tabular}} 
& The Russian armed forces are not striking at the  & Russian forces continue strikes in multiple cities $[...]$.  \\
&  cities of Ukraine; they are not threatening the  & This is premeditated mass murder and must be responded  \\
&  civilian population. & to as such. \\ \bottomrule
    \end{tabular}}

\end{table*}


To identify false and unsubstantiated claims that circulated online during the same period, we rely on the Russia-Ukraine ConflictMisinfo Dashboard\footnote{\url{www.shorturl.at/puN37}}, which provides a collection of true and false claims and rumors verified by fact-checking outlets such as \emph{USA Today} and \emph{Snopes}. Specifically, we collect $83$ English false claims that were verified in the period 22nd February - March 1st. 

In addition, we leverage an existing dataset\footnote{Reference omitted for blind review.} of tweets matching over 30 conflict-related keywords in English, Russian and Ukrainian language, which were identified through a snowball sampling approach, collected via Twitter's Filter v1.1 Streaming  API\footnote{\url{https://developer.twitter.com/en/docs/twitter-api/v1}}. In particular, we focus on English-language tweets shared during the initial weeks of the invasion, i.e., from February 22nd, 2022 to March 8th, 2022. Note that we consider the tweets posted until one week after the above-mentioned false claims were verified (March 1st) so as to capture their propagation on Twitter. Overall, in this observation period, the collected dataset contains more than 2M English tweets with original content, i.e., original tweets, replies, and quotes. We purposely exclude retweets as their textual content is exactly the same. 

\subsection{Data Annotation}
\label{sec:dataset}
Given the collected fact-checked information, our goal is to find tweets reporting such verified claims.
However, as manually annotating the whole corpus of tweets in our dataset is not a suitable solution, we deploy a machine learning-based annotation strategy to maximize the likelihood of finding tweets related to one of the claims under analysis. 

Specifically, we first use the RoBERTa transformer \cite{DBLP:journals/corr/abs-1907-11692} to extract the vector embeddings of both claims and tweets, and then we compute the cosine similarity between each claim and tweet in our data, retaining the top-100 most similar tweets for each claim.  Consequently, we end up with 8,300 unique tweet-claim pairs that were inspected via a \emph{manual} labeling process. It is worth noting that the choice of the RoBERTa transformer depends on the higher similarity scores provided by this model with respect to other transformers (\emph{ms-marco-MiniLM-L-4-v2} and \emph{quora-roberta-base}), which allows us to maximize the chance of finding a matching pair.  


Next, we annotate the above-mentioned tweet-claim pairs considering a strict definition of relevance, i.e., the tweet discusses the claim if it explicitly mentions the same entities and events reported in the fact-checked information. Table \ref{tab:examples} shows some examples of false claims and matched tweets, which highlight how the matched tweet can discuss a claim without expressing any stance (examples no. 1 and 2) or can support or refute the claim (examples no. 3 and 4, respectively). Finally, it is worth noting that the number of unique tweets is different from the number of pairs because a tweet can be related to multiple claims. In particular, we find 5,872 unique tweets in the ranking of the 83 claims. 



\begin{figure}[t]
  \centering
  \includegraphics[width=\columnwidth]{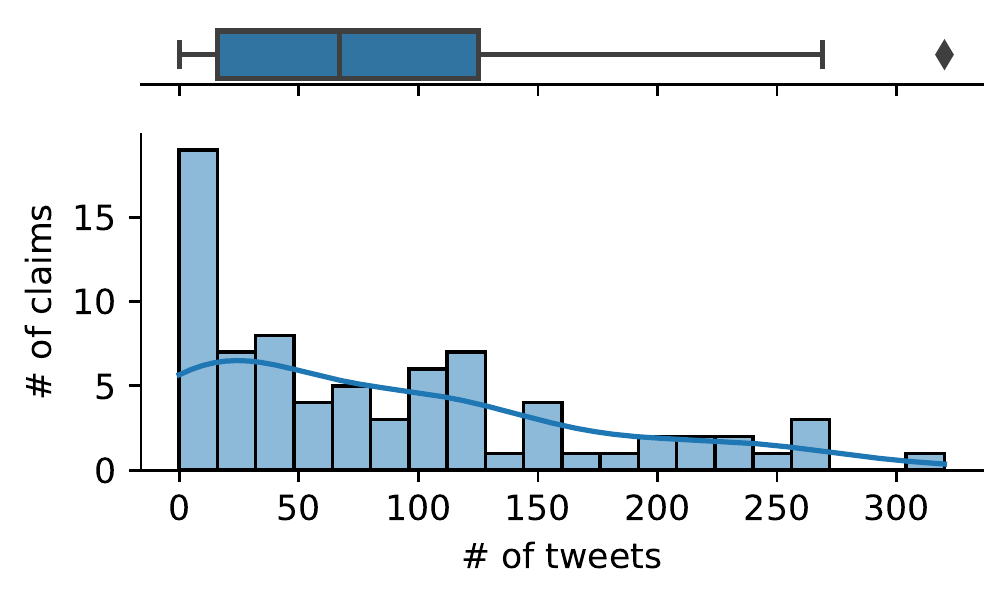}
  \caption{Distribution of the number of tweets with respect to the number of claims }
  \label{fig:dist}
\end{figure}

Our manual annotation results in 2,359 (out of 5,872 -- $40.2\%$) tweets associated with at least one claim. Figure \ref{fig:dist} shows the distribution of the number of tweets with respect to the number of claims: most of the tweets are related to less than five claims and only 13 tweets discuss more than 10 claims. Overall, each claim has at least one matching tweet and the most matched claim has 100 related tweets. Conversely, we find 3,513 (out of 5,872 -- $59.8\%$) tweets that do not report any claim.

\begin{figure}[t]
  \centering
  \includegraphics[width=\columnwidth]{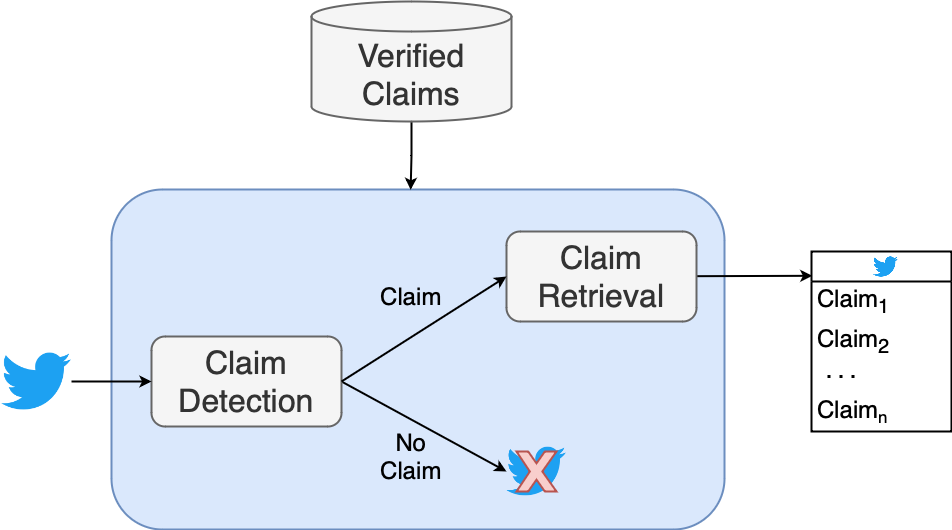}
  \caption{Our proposed methodological framework: the \emph{claim detection} model detects whether the input tweet reports a fact-checked (false) claim. If a claim is detected, the \emph{claim retrieval} model retrieves the most relevant claims (within the corpus of ``Verified Claims'') related to the tweet. }
  \label{fig:model_design}
\end{figure}

\subsection{Our Framework}

Figure \ref{fig:model_design} describes the two-step pipeline proposed in our methodological framework. The first step is characterized by a \emph{claim detection} model, which verifies whether an input tweet discusses a (false) claim included in a predetermined set of already verified claims. Next, if the input tweet actually discusses a false claim in the \emph{Verified Claims} corpus, a \emph{claim retrieval} model ranks the claims in the corpus according to their relevance with the input tweet. It is worth noting that, even if we target false claims, our methodology does not depend on the claim's truthfulness and can be adopted to retrieve tweets matching any claim associated with a specific topic. 



\subsubsection{Claim Detection}
\label{sec:cd}


Overall, we frame the claim detection problem as a binary classification task: given an input tweet $t$, the objective is to establish whether $t$ contains a false claim among the ones collected in the corpus of \emph{Verified Claims}. Therefore, we leverage BERT-based language models \cite{DBLP:journals/corr/abs-1810-04805}, which have proven their effectiveness in several text classification tasks \cite{DBLP:journals/corr/abs-2106-04554} due to the extensive knowledge they acquire during the pre-training process. In particular, we fine-tune the \emph{BERT-base-cased} with the Huggingface library\footnote{https://huggingface.co} for five epochs using Adam optimizer, categorical cross-entropy as a loss function, and 20 tweets as batch size.

In addition, assuming a realistic scenario where only a minority of tweets report specific factual events related to the war \cite{DBLP:journals/corr/abs-2008-08854} (i.e., 2.3k out of 5.9k tweets identified by our annotation procedure), we evaluate our model by performing random oversampling of the minority class (i.e., tweets with a specific claim). In other words, we randomly replicate tweets containing a claim so as to have the same number of tweets belonging to the positive and negative classes. 

\subsubsection{Claim Retrieval}
Following the cascade architecture of our framework (see Fig. \ref{fig:model_design}), we 
are now interested in finding the most relevant claims associated with the input tweets that report a claim, as determined with the \emph{claim detection} model.\footnote{Note that if the claim detection model assesses the input tweet to be unrelated to any claim, the input tweet is automatically discarded and not considered in the claim retrieval model.}
Formally, we leverage the information retrieval formulation related to the task of identifying fact-checked claims \cite{shaar-etal-2020-known}: given an input tweet $t$ containing a claim, rank a set of verified (false) claims $\{c_1, c_2, \cdots, c_n\}$ based on their relevance to $t$. 
 
In particular, we consider the tweet-claim pairs as input of the system and use the similarity score to rank the verified claims. However, performing the supervised training process, we need to build the negative class by finding instances of unrelated tweet-claim pairs. This problem is common when addressing information retrieval tasks \cite{10.1145/3486250}, and several (negative) sampling strategies have been evaluated across different ranking tasks \cite{li-etal-2019-sampling,DBLP:journals/corr/abs-2007-00808}. Among the others, we find that random negative sampling yields good results in our use scenario, i.e., for each tweet we randomly select 10 unrelated claims.

Next, as for the claim detection model, we leverage a transformer-based architecture. Indeed, when optimized for ranking tasks, pre-trained language models have proven their effectiveness in a variety of information retrieval tasks, including semantic search \cite{DBLP:journals/corr/abs-1908-02451} and semantic textual similarity \cite{DBLP:journals/corr/abs-1908-10084}. Specifically, we finetune the \emph{stsb-roberta-base} cross-encoder within the SBERT library\footnote{https://www.sbert.net}. 

We train the model for three epochs using categorical cross-entropy loss and 16 tweet-claim pairs as batch size. 



\section{Experiments}
This section presents the experiments we have performed to evaluate our framework. All experiments are conducted on a machine equipped with CPU Intel Xeon-4116, RAM 32 GB, and one NVIDIA A100. 

\subsection{Evaluation Data and Metrics}
In this section, we describe both the data sets and metrics used to evaluate our methodological framework. We distinguish these for the claim detection and claim retrieval models, given the different prediction goals of the two tasks.
For the claim detection task, we consider the annotated tweets described beforehand. Our dataset includes 2,359 (out of 5,872 -- 40,2\%) tweets reporting at least one (false) claim and 3,513  (out of 5,872 -- 59,8\%) tweets that do not incorporate any claim. Given the classification task, we measure the performance of our models considering binary classification metrics such as precision, recall, f-score, and accuracy. 

For the claim retrieval task, we consider the 2,359 tweets related to the 83 claims gathered from the Russia-Ukraine ConflictMisinfo Dashboard. We build up a dataset of 40,007 
tweet-claim pairs as previously discussed, consisting of 3,637 \emph{positive} tweet-claim pairs and 36,370 \emph{negative} tweet-claim pairs.
Consistently with previous work \cite{vo2020facts,shaar-etal-2020-known}, we evaluate the performance of the claim retrieval model with the HitRatio@K, i.e., whether the correct claim (i.e., the claim matching the tweet) is among the top-k results of the ranking. It is worth noting that, since most of the tweets match only one claim, HitRatio@K is almost equal to Recall@K \cite{vo2020facts}. On the one hand, from the end-user's perspective, the HitRatio on lower values of $k$ (e.g. $k \in \{1,3,5\}$) might be indicative of the system utility in easing manual fact-checkers works, i.e., experts would spot in real-time if the top-ranked results are relevant to the input claim. On the other hand, the HitRatio on higher values of $k$ (e.g. $k \in \{10,20, 50\}$) should be considered in offline settings and/or in an automated fact-checking pipeline where results should be further processed as evidence for the veracity prediction. 

\subsection{Experimental Protocol}

\begin{table}
    \centering

    \caption{Claim detection: performance with and without random oversampling}
\label{tab:oversampling}

    \scalebox{1.}{
    \begin{tabular}{r|cc}\toprule
    
  \textbf{Metric} & \textbf{No Oversampling} & \textbf{With Oversampling}  \\ \midrule

Precision & 79.90\% & \textbf{81.39\%} \\
Recall    & 80.19\% & \textbf{80.24\%} \\
F1-score & 79.35\% & \textbf{80.57\%} \\
Accuracy  & 80.11\% & \textbf{81.59\%} \\ \bottomrule
\end{tabular}}

\end{table}

\begin{table*}
    \centering

\caption{Claim detection: performance comparison per class, and their 95\% confidence interval, between TF-IDF baseline and our approach (bold indicates best on average, $^*$ indicates statistical significance ($p<0.01$)}
\label{tab:per_class}

    \scalebox{1.}{
    \begin{tabular}{r|cc|cc}\toprule
    
  \textbf{Metric} & \multicolumn{2}{c}{Claim (Positive class)} & \multicolumn{2}{c}{No Claim (Negative class)}  \\  \cmidrule{2-5}

  & TF-IDF & Ours & TF-IDF & Ours \\ \midrule

Precision  & $67.88\% \pm 1.57\% $ & $\mathbf{79.75\%  \pm 3.05\%}^*$  & $80.97\% \pm 2.92\% $ & $\mathbf{83.02\%   \pm 1.87\%}$    \\
Recall     & $73.01\% \pm 4.99\% $ & $\mathbf{73.31\%  \pm 2.09\%}$  & $76.84\% \pm 0.81\% $ & $\mathbf{87.17\%   \pm 3.85\%}^*$    \\
F1-score   & $70.33\% \pm 3.05\% $ & $\mathbf{76.17\%  \pm 1.71\%}^*$  & $78.84\% \pm 1.40\% $ & $\mathbf{84.97\%   \pm 1.39\%}^*$  \\ \bottomrule

\end{tabular}}

\end{table*}

 For each task, we perform a 5-fold cross-validation to evaluate the performance of our models. In particular, we consider two evaluation settings:

 \begin{itemize}
     \item Leave Tweet Out (LTO) Assessment: for both the claim detection and retrieval tasks, we ensure that the tweets in the training, validation, and testing sets do not overlap.  
     \item Leave Claim Out (LCO) Assessment: for the claim detection task, we ensure that the tweets in the training, validation, and testing sets match different claims. For the claim retrieval task, we ensure that the claims within the tweet-claim pairs in the training, validation, and testing sets do not overlap.
 \end{itemize}

The LTO assessment refers to the classical evaluation of supervised machine learning approaches and is consistent with experiments from previous works \cite{shaar-etal-2020-known,vo2020facts,DBLP:journals/corr/abs-2109-12987}. The LCO assessment focuses on the generalization of the models to unseen claims and is close to a real-world scenario. Indeed, when deployed "in the wild", both claim detection and retrieval should be performed on tweets and claims that the models have not seen during the training process. 

When evaluating the performance of the claim detection task, we compare our BERT-based model to a TF-IDF baseline \cite{10.1145/3412869}, also assessing the effectiveness of the oversampling strategy detailed in Section \ref{sec:cd}. For the claim retrieval task, we compare the ranking performance of our BERT-based cross-encoder with respect to Sentence-BERT \cite{DBLP:journals/corr/abs-1908-10084}. 

\subsection{Results}

\subsubsection{Claim Detection}

Table \ref{tab:oversampling} shows classification performance with and without the random oversampling strategy. In particular, we can notice that oversampling the minority class improves the predictive power of the claim detection model for every classification metric. 

Table \ref{tab:per_class} shows the performance marginalized for each class of the \emph{claim detection} task. In particular, our BERT-based model outperforms the TF-IDF baseline for both classes and all metrics. Overall, we can observe that performance is better on the negative class, i.e., the system has higher precision and recall in the detection of tweets that do not report any claim. We do not report the accuracy metric as it is equivalent to the recall computed on the single class. 

Finally, Table \ref{tab:lto_lco} shows the aggregated performance under the LTO and LCO evaluation settings. We can observe that LCO evaluation is more challenging since the model has to deal with claims that it has never seen during the training process. 

\begin{table}
    \centering

    \caption{Claim detection: LTO and LCO assessment}
\label{tab:lto_lco}

    \scalebox{1.}{
    \begin{tabular}{r|cc} 
    & \multicolumn{2}{c}{Settings} \\ 
    
    \toprule
    
  \textbf{Metric} & \textbf{LTO} & \textbf{LCO}  \\ \midrule

Precision & $81.39\% \pm 1.14\% $ & $78.07\% \pm 1.48\% $ \\
Recall    & $80.24\% \pm 1.36\% $ & $77.58\% \pm 1.67\% $ \\
F1-score  & $80.57\% \pm 1.38\% $ & $77.63\% \pm 1.60\% $ \\
Accuracy  & $81.59\% \pm 1.42\% $ & $78.03\% \pm 1.56\% $ \\ \bottomrule
\end{tabular}}

\end{table}

\subsubsection{Claim Retrieval}
Table \ref{tab:retrieval} shows the ranking performance of our claim retrieval model in comparison with sentence-BERT \cite{DBLP:journals/corr/abs-1908-10084}. In both evaluation settings (i.e., LTO and LCO), our model outperforms the baseline, and the performance gap is even larger when considering the top positions in the ranking ($k \in \{1,3\}$). This result highlights that our system can actually ease fact-checkers' work because it can spot tweets that discuss false claims that were already verified. In addition, we can notice again that the performance under LTO settings is slightly better than the ones under LCO settings, even if the gap is smaller with respect to what we found in the claim detection task. 

\begin{table*}
    \centering

    \caption{Claim retrieval: performance comparison, and their 95\% confidence interval, between the sentence-BERT baseline and our approach (bold indicates best on average, $^*$ indicates statistical significance ($p<0.01$)}

    \label{tab:retrieval}
    
\resizebox{\textwidth}{!}{  
    \begin{tabular}{rcccccc}\toprule
    
Setting & Model  &  \multicolumn{5}{c}{HitRatio{$@k$}} \\
 & &  $k=1$ & $k=3$ & $k=5$ & $k=10$& $k=20$  \\ \midrule

\multirow{2}{*}{\begin{tabular}[c]{c}LTO\end{tabular}} &
Sentence-BERT   & $85.25\% \pm 2.07\% $ &  $94.87\% \pm 1.69\% $ & $97.24\% \pm 0.96\% $ & $98.77\% \pm 0.43\%1 $ & $99.27\% \pm 0.39\% $   \\ 
& Ours & $\mathbf{86.05\% \pm 0.95\% }$ &  $\mathbf{96.35\% \pm 0.71\% }^*$ & $\mathbf{98.04\% \pm 0.57\% }^*$ & $\mathbf{99.27\% \pm 0.36\%} $ & $\mathbf{99.78\% \pm 0.11\%}^* $   \\ \midrule

\multirow{2}{*}{\begin{tabular}[c]{c}LCO\end{tabular}} &
Sentence-BERT   & $77.60\% \pm 0.1196 $ &  $95.68\% \pm 6.74\% $ & $98.01\% \pm 3.66\% $ & $99.63\% \pm 0.66\% $ & $99.78\% \pm 0.00\% $   \\ 
& Ours & $\mathbf{82.25\% \pm 10.81\%}^* $ &  $\mathbf{96.42\% \pm 2.59\%} $ & $\mathbf{98.26\% \pm 1.45\%} $ & $\mathbf{99.88\% \pm 0.02\%} $ & $\mathbf{99.96\% \pm 0.00\%} $  \\ \bottomrule

    \end{tabular}
    }

\end{table*}

\subsubsection{Evaluation in the wild}

Finally, we apply our claim detection and retrieval models on the whole Twitter dataset to assess how our methodology performs in a real scenario.  In this case, we do not have any annotation for the LTO and LCO settings, and thus we cannot perform a formal quantitative analysis. However, we randomly sample 100 (resp. 100) tweets assessed as reporting a claim (resp. not reporting any claim) by the claim detection model. Then, we manually verify whether the models make the correct prediction, i.e., whether the tweet discusses a false claim and, if so, to which false claim it was related.


Out of the 100 tweets that are predicted by the \emph{claim detection} model as not related to any (false) claim, we find 12 misclassified instances, i.e., tweets that are actually related to a false claim. This result confirms that the \emph{claim detection} model has a high precision ($>80\%$) on the negative class (see Table \ref{tab:per_class} for further details). 

On the contrary, out of the 100 tweets that are predicted by the \emph{claim detection} model as discussing a (false) claim, we find 64 false positives, i.e., tweets do not discuss any false claim but generally report information about the Ukraine-Russia war. This result confirms the performance differences between the positive and negative classes (see Table \ref{tab:per_class}) of the \emph{claim detection} model, and probably depends on data imbalance, i.e., the number of tweets with no claim is higher than the number of tweets reporting a claim. 


In addition, we apply the \emph{claim retrieval} model for the 36 tweets that were actually related to some false claims. In particular, we retrieve the top-3 relevant false claims and find that the correct matching claim was retrieved 34 times (out of 36). 


Overall, these results confirm the promising performance of the \emph{claim retrieval} model but highlight the limitations of the \emph{claim detection} model, which overestimates the number of tweets discussing false claims.

\section{Conclusions and Future Work}

We presented a methodological framework to detect over 80 false and unsubstantiated claims that were shared on Twitter during the first week of the conflict. Our framework first performs \emph{claim detection} to identify whether an input tweet contains a (false) claim or not. Then, assuming the input tweet reports a claim, the system performs \emph{claim retrieval} to rank a set of already-verified false claims according to their relevance with the input tweet. When fine-tuning modern BERT-based models, our methodology achieves promising performance to automate both tasks. Indeed, our models show good generalization capabilities, i.e., they reach good performance even when the claims that need to be detected were never seen during the training process. 

Despite the promising performance, we highlight some limitations of our approach. First, we considered a limited set of false claims because of the short observation window, i.e., the first week of the invasion. However, when focusing on a longer time period, the number of verified claims increases as well as what was considered false at the beginning of the conflict could become true afterward (e.g., NATO's members providing Ukraine with military weapons). Second, when collecting tweets discussing false claims, we did not consider their stance and implicitly assumed 
they support or divulge the claims \cite{youtubetweetpair}. However, we actually find some tweets that reported (false) claims debating their veracity.

There is a number of avenues to explore in future research. First, we aim to extend our analysis to multimodal data, i.e., images and videos that were shared within the tweets could help our system to improve its classification and ranking performance. Second, we plan to apply our methodology ``into the wild'' to investigate the diffusion of the most shared (false) claims on Twitter and unveil the communities that support or are most susceptible to false narratives. Third, we will deploy our framework within a fact-checking pipeline and assess the extent to which such a system could improve manual fact-checking by allowing debunkers to focus on brand-new claims and ignoring similar claims that have been already verified.  

\section{Acknowledgments}
Work supported in part by
DARPA (contract \#HR001121C0169) and PRIN grant HOPE (FP6, Italian Ministry of Education).

\bibliographystyle{ACM-Reference-Format}
\bibliography{sample-base.bib}


\begin{thebibliography}{35}


\ifx \showCODEN    \undefined \def \showCODEN     #1{\unskip}     \fi
\ifx \showDOI      \undefined \def \showDOI       #1{#1}\fi
\ifx \showISBNx    \undefined \def \showISBNx     #1{\unskip}     \fi
\ifx \showISBNxiii \undefined \def \showISBNxiii  #1{\unskip}     \fi
\ifx \showISSN     \undefined \def \showISSN      #1{\unskip}     \fi
\ifx \showLCCN     \undefined \def \showLCCN      #1{\unskip}     \fi
\ifx \shownote     \undefined \def \shownote      #1{#1}          \fi
\ifx \showarticletitle \undefined \def \showarticletitle #1{#1}   \fi
\ifx \showURL      \undefined \def \showURL       {\relax}        \fi
\providecommand\bibfield[2]{#2}
\providecommand\bibinfo[2]{#2}
\providecommand\natexlab[1]{#1}
\providecommand\showeprint[2][]{arXiv:#2}

\bibitem[Allein and Moens(2020)]%
        {DBLP:journals/corr/abs-2008-08854}
\bibfield{author}{\bibinfo{person}{Liesbeth Allein} {and}
  \bibinfo{person}{Marie{-}Francine Moens}.} \bibinfo{year}{2020}\natexlab{}.
\newblock \showarticletitle{Checkworthiness in Automatic Claim Detection
  Models: Definitions and Analysis of Datasets}.
\newblock \bibinfo{journal}{\emph{CoRR}}  \bibinfo{volume}{abs/2008.08854}
  (\bibinfo{year}{2020}).
\newblock
\showeprint[arXiv]{2008.08854}
\urldef\tempurl%
\url{https://arxiv.org/abs/2008.08854}
\showURL{%
\tempurl}


\bibitem[Atanasova et~al\mbox{.}(2019)]%
        {10.1145/3297722}
\bibfield{author}{\bibinfo{person}{Pepa Atanasova}, \bibinfo{person}{Preslav
  Nakov}, \bibinfo{person}{Llu\'{\i}s M\`{a}rquez}, \bibinfo{person}{Alberto
  Barr\'{o}n-Cede\~{n}o}, \bibinfo{person}{Georgi Karadzhov},
  \bibinfo{person}{Tsvetomila Mihaylova}, \bibinfo{person}{Mitra Mohtarami},
  {and} \bibinfo{person}{James Glass}.} \bibinfo{year}{2019}\natexlab{}.
\newblock \showarticletitle{Automatic Fact-Checking Using Context and Discourse
  Information}.
\newblock \bibinfo{journal}{\emph{J. Data and Information Quality}}
  \bibinfo{volume}{11}, \bibinfo{number}{3}, Article \bibinfo{articleno}{12}
  (\bibinfo{date}{may} \bibinfo{year}{2019}), \bibinfo{numpages}{27}~pages.
\newblock
\showISSN{1936-1955}
\urldef\tempurl%
\url{https://doi.org/10.1145/3297722}
\showDOI{\tempurl}


\bibitem[Badawy et~al\mbox{.}(2018)]%
        {badawy2018analyzing}
\bibfield{author}{\bibinfo{person}{Adam Badawy}, \bibinfo{person}{Emilio
  Ferrara}, {and} \bibinfo{person}{Kristina Lerman}.}
  \bibinfo{year}{2018}\natexlab{}.
\newblock \showarticletitle{Analyzing the digital traces of political
  manipulation: The 2016 Russian interference Twitter campaign}. In
  \bibinfo{booktitle}{\emph{2018 IEEE/ACM international conference on advances
  in social networks analysis and mining (ASONAM)}}. IEEE,
  \bibinfo{pages}{258--265}.
\newblock


\bibitem[Chen et~al\mbox{.}(2022)]%
        {Chen_2022}
\bibfield{author}{\bibinfo{person}{Jiangui Chen}, \bibinfo{person}{Ruqing
  Zhang}, \bibinfo{person}{Jiafeng Guo}, \bibinfo{person}{Yixing Fan}, {and}
  \bibinfo{person}{Xueqi Cheng}.} \bibinfo{year}{2022}\natexlab{}.
\newblock \showarticletitle{{GERE}: Generative Evidence Retrieval for Fact
  Verification}. In \bibinfo{booktitle}{\emph{Proceedings of the 45th
  International {ACM} {SIGIR} Conference on Research and Development in
  Information Retrieval}}. \bibinfo{publisher}{{ACM}}.
\newblock
\urldef\tempurl%
\url{https://doi.org/10.1145/3477495.3531827}
\showDOI{\tempurl}


\bibitem[Devlin et~al\mbox{.}(2018)]%
        {DBLP:journals/corr/abs-1810-04805}
\bibfield{author}{\bibinfo{person}{Jacob Devlin}, \bibinfo{person}{Ming{-}Wei
  Chang}, \bibinfo{person}{Kenton Lee}, {and} \bibinfo{person}{Kristina
  Toutanova}.} \bibinfo{year}{2018}\natexlab{}.
\newblock \showarticletitle{{BERT:} Pre-training of Deep Bidirectional
  Transformers for Language Understanding}.
\newblock \bibinfo{journal}{\emph{CoRR}}  \bibinfo{volume}{abs/1810.04805}
  (\bibinfo{year}{2018}).
\newblock
\showeprint[arXiv]{1810.04805}
\urldef\tempurl%
\url{http://arxiv.org/abs/1810.04805}
\showURL{%
\tempurl}


\bibitem[Guo et~al\mbox{.}(2022a)]%
        {10.1145/3486250}
\bibfield{author}{\bibinfo{person}{Jiafeng Guo}, \bibinfo{person}{Yinqiong
  Cai}, \bibinfo{person}{Yixing Fan}, \bibinfo{person}{Fei Sun},
  \bibinfo{person}{Ruqing Zhang}, {and} \bibinfo{person}{Xueqi Cheng}.}
  \bibinfo{year}{2022}\natexlab{a}.
\newblock \showarticletitle{Semantic Models for the First-Stage Retrieval: A
  Comprehensive Review}.
\newblock \bibinfo{journal}{\emph{ACM Trans. Inf. Syst.}} \bibinfo{volume}{40},
  \bibinfo{number}{4}, Article \bibinfo{articleno}{66} (\bibinfo{date}{mar}
  \bibinfo{year}{2022}), \bibinfo{numpages}{42}~pages.
\newblock
\showISSN{1046-8188}
\urldef\tempurl%
\url{https://doi.org/10.1145/3486250}
\showDOI{\tempurl}


\bibitem[Guo et~al\mbox{.}(2022b)]%
        {DBLP:journals/corr/abs-2108-11896}
\bibfield{author}{\bibinfo{person}{Zhijiang Guo}, \bibinfo{person}{Michael
  Schlichtkrull}, {and} \bibinfo{person}{Andreas Vlachos}.}
  \bibinfo{year}{2022}\natexlab{b}.
\newblock \showarticletitle{A Survey on Automated Fact-Checking}.
\newblock \bibinfo{journal}{\emph{Transactions of the Association for
  Computational Linguistics}}  \bibinfo{volume}{10} (\bibinfo{year}{2022}),
  \bibinfo{pages}{178--206}.
\newblock
\urldef\tempurl%
\url{https://doi.org/10.1162/tacl_a_00454}
\showDOI{\tempurl}


\bibitem[Gupta et~al\mbox{.}(2021)]%
        {DBLP:journals/corr/abs-2101-11891}
\bibfield{author}{\bibinfo{person}{Shreya Gupta}, \bibinfo{person}{Parantak
  Singh}, \bibinfo{person}{Megha Sundriyal}, \bibinfo{person}{Md.~Shad Akhtar},
  {and} \bibinfo{person}{Tanmoy Chakraborty}.} \bibinfo{year}{2021}\natexlab{}.
\newblock \showarticletitle{{LESA:} Linguistic Encapsulation and Semantic
  Amalgamation Based Generalised Claim Detection from Online Content}.
\newblock \bibinfo{journal}{\emph{CoRR}}  \bibinfo{volume}{abs/2101.11891}
  (\bibinfo{year}{2021}).
\newblock
\showeprint[arXiv]{2101.11891}
\urldef\tempurl%
\url{https://arxiv.org/abs/2101.11891}
\showURL{%
\tempurl}


\bibitem[Hanley et~al\mbox{.}(2022a)]%
        {hanley2022special}
\bibfield{author}{\bibinfo{person}{Hans~WA Hanley}, \bibinfo{person}{Deepak
  Kumar}, {and} \bibinfo{person}{Zakir Durumeric}.}
  \bibinfo{year}{2022}\natexlab{a}.
\newblock \showarticletitle{" A Special Operation": A Quantitative Approach to
  Dissecting and Comparing Different Media Ecosystems' Coverage of the
  Russo-Ukrainian War}.
\newblock \bibinfo{journal}{\emph{arXiv preprint arXiv:2210.03016}}
  (\bibinfo{year}{2022}).
\newblock


\bibitem[Hanley et~al\mbox{.}(2022b)]%
        {hanley2022happenstance}
\bibfield{author}{\bibinfo{person}{Hans~WA Hanley}, \bibinfo{person}{Deepak
  Kumar}, {and} \bibinfo{person}{Zakir Durumeric}.}
  \bibinfo{year}{2022}\natexlab{b}.
\newblock \showarticletitle{Happenstance: Utilizing Semantic Search to Track
  Russian State Media Narratives about the Russo-Ukrainian War On Reddit}.
\newblock \bibinfo{journal}{\emph{arXiv preprint arXiv:2205.14484}}
  (\bibinfo{year}{2022}).
\newblock


\bibitem[Hassan et~al\mbox{.}(2015)]%
        {10.1145/2806416.2806652}
\bibfield{author}{\bibinfo{person}{Naeemul Hassan}, \bibinfo{person}{Chengkai
  Li}, {and} \bibinfo{person}{Mark Tremayne}.} \bibinfo{year}{2015}\natexlab{}.
\newblock \showarticletitle{Detecting Check-Worthy Factual Claims in
  Presidential Debates}. In \bibinfo{booktitle}{\emph{Proceedings of the 24th
  ACM International on Conference on Information and Knowledge Management}}
  (Melbourne, Australia) \emph{(\bibinfo{series}{CIKM '15})}.
  \bibinfo{publisher}{Association for Computing Machinery},
  \bibinfo{address}{New York, NY, USA}, \bibinfo{pages}{1835–1838}.
\newblock
\showISBNx{9781450337946}
\urldef\tempurl%
\url{https://doi.org/10.1145/2806416.2806652}
\showDOI{\tempurl}


\bibitem[Konstantinovskiy et~al\mbox{.}(2021)]%
        {10.1145/3412869}
\bibfield{author}{\bibinfo{person}{Lev Konstantinovskiy},
  \bibinfo{person}{Oliver Price}, \bibinfo{person}{Mevan Babakar}, {and}
  \bibinfo{person}{Arkaitz Zubiaga}.} \bibinfo{year}{2021}\natexlab{}.
\newblock \showarticletitle{Toward Automated Factchecking: Developing an
  Annotation Schema and Benchmark for Consistent Automated Claim Detection}.
\newblock \bibinfo{journal}{\emph{Digital Threats}} \bibinfo{volume}{2},
  \bibinfo{number}{2}, Article \bibinfo{articleno}{14} (\bibinfo{date}{apr}
  \bibinfo{year}{2021}), \bibinfo{numpages}{16}~pages.
\newblock
\showISSN{2692-1626}
\urldef\tempurl%
\url{https://doi.org/10.1145/3412869}
\showDOI{\tempurl}


\bibitem[Li et~al\mbox{.}(2019)]%
        {li-etal-2019-sampling}
\bibfield{author}{\bibinfo{person}{Jia Li}, \bibinfo{person}{Chongyang Tao},
  \bibinfo{person}{Wei Wu}, \bibinfo{person}{Yansong Feng},
  \bibinfo{person}{Dongyan Zhao}, {and} \bibinfo{person}{Rui Yan}.}
  \bibinfo{year}{2019}\natexlab{}.
\newblock \showarticletitle{Sampling Matters! An Empirical Study of Negative
  Sampling Strategies for Learning of Matching Models in Retrieval-based
  Dialogue Systems}. In \bibinfo{booktitle}{\emph{Proceedings of the 2019
  Conference on Empirical Methods in Natural Language Processing and the 9th
  International Joint Conference on Natural Language Processing
  (EMNLP-IJCNLP)}}. \bibinfo{publisher}{Association for Computational
  Linguistics}, \bibinfo{address}{Hong Kong, China},
  \bibinfo{pages}{1291--1296}.
\newblock
\urldef\tempurl%
\url{https://doi.org/10.18653/v1/D19-1128}
\showDOI{\tempurl}


\bibitem[Lin et~al\mbox{.}(2021)]%
        {DBLP:journals/corr/abs-2106-04554}
\bibfield{author}{\bibinfo{person}{Tianyang Lin}, \bibinfo{person}{Yuxin Wang},
  \bibinfo{person}{Xiangyang Liu}, {and} \bibinfo{person}{Xipeng Qiu}.}
  \bibinfo{year}{2021}\natexlab{}.
\newblock \showarticletitle{A Survey of Transformers}.
\newblock \bibinfo{journal}{\emph{CoRR}}  \bibinfo{volume}{abs/2106.04554}
  (\bibinfo{year}{2021}).
\newblock
\showeprint[arXiv]{2106.04554}
\urldef\tempurl%
\url{https://arxiv.org/abs/2106.04554}
\showURL{%
\tempurl}


\bibitem[Liu et~al\mbox{.}(2019)]%
        {DBLP:journals/corr/abs-1907-11692}
\bibfield{author}{\bibinfo{person}{Yinhan Liu}, \bibinfo{person}{Myle Ott},
  \bibinfo{person}{Naman Goyal}, \bibinfo{person}{Jingfei Du},
  \bibinfo{person}{Mandar Joshi}, \bibinfo{person}{Danqi Chen},
  \bibinfo{person}{Omer Levy}, \bibinfo{person}{Mike Lewis},
  \bibinfo{person}{Luke Zettlemoyer}, {and} \bibinfo{person}{Veselin
  Stoyanov}.} \bibinfo{year}{2019}\natexlab{}.
\newblock \showarticletitle{RoBERTa: {A} Robustly Optimized {BERT} Pretraining
  Approach}.
\newblock \bibinfo{journal}{\emph{CoRR}}  \bibinfo{volume}{abs/1907.11692}
  (\bibinfo{year}{2019}).
\newblock
\showeprint[arxiv]{1907.11692}
\urldef\tempurl%
\url{http://arxiv.org/abs/1907.11692}
\showURL{%
\tempurl}


\bibitem[Luceri et~al\mbox{.}(2020)]%
        {luceri2020detecting}
\bibfield{author}{\bibinfo{person}{Luca Luceri}, \bibinfo{person}{Silvia
  Giordano}, {and} \bibinfo{person}{Emilio Ferrara}.}
  \bibinfo{year}{2020}\natexlab{}.
\newblock \showarticletitle{Detecting troll behavior via inverse reinforcement
  learning: A case study of Russian trolls in the 2016 US election}. In
  \bibinfo{booktitle}{\emph{Proceedings of the International AAAI Conference on
  Web and Social Media}}, Vol.~\bibinfo{volume}{14}. \bibinfo{pages}{417--427}.
\newblock


\bibitem[Mansour et~al\mbox{.}(2022)]%
        {10.1007/978-3-030-99736-6_25}
\bibfield{author}{\bibinfo{person}{Watheq Mansour}, \bibinfo{person}{Tamer
  Elsayed}, {and} \bibinfo{person}{Abdulaziz Al-Ali}.}
  \bibinfo{year}{2022}\natexlab{}.
\newblock \showarticletitle{Did I See It Before? Detecting Previously-Checked
  Claims over Twitter}. In \bibinfo{booktitle}{\emph{Advances in Information
  Retrieval: 44th European Conference on IR Research, ECIR 2022, Stavanger,
  Norway, April 10–14, 2022, Proceedings, Part I}} (Stavanger, Norway).
  \bibinfo{publisher}{Springer-Verlag}, \bibinfo{address}{Berlin, Heidelberg},
  \bibinfo{pages}{367–381}.
\newblock
\showISBNx{978-3-030-99735-9}
\urldef\tempurl%
\url{https://doi.org/10.1007/978-3-030-99736-6_25}
\showDOI{\tempurl}


\bibitem[Micallef et~al\mbox{.}(2022)]%
        {youtubetweetpair}
\bibfield{author}{\bibinfo{person}{Nicholas Micallef}, \bibinfo{person}{Marcelo
  Sandoval-Castañeda}, \bibinfo{person}{Adi Cohen}, \bibinfo{person}{Mustaque
  Ahamad}, \bibinfo{person}{Srijan Kumar}, {and} \bibinfo{person}{Nasir
  Memon}.} \bibinfo{year}{2022}\natexlab{}.
\newblock \showarticletitle{Cross-Platform Multimodal Misinformation: Taxonomy,
  Characteristics and Detection for Textual Posts and Videos}.
\newblock \bibinfo{journal}{\emph{Proceedings of the International AAAI
  Conference on Web and Social Media}} \bibinfo{volume}{16},
  \bibinfo{number}{1} (\bibinfo{date}{May} \bibinfo{year}{2022}),
  \bibinfo{pages}{651--662}.
\newblock
\urldef\tempurl%
\url{https://doi.org/10.1609/icwsm.v16i1.19323}
\showDOI{\tempurl}


\bibitem[Nakov et~al\mbox{.}(2021a)]%
        {DBLP:journals/corr/abs-2103-07769}
\bibfield{author}{\bibinfo{person}{Preslav Nakov}, \bibinfo{person}{David P.~A.
  Corney}, \bibinfo{person}{Maram Hasanain}, \bibinfo{person}{Firoj Alam},
  \bibinfo{person}{Tamer Elsayed}, \bibinfo{person}{Alberto
  Barr{\'{o}}n{-}Cede{\~{n}}o}, \bibinfo{person}{Paolo Papotti},
  \bibinfo{person}{Shaden Shaar}, {and} \bibinfo{person}{Giovanni Da~San
  Martino}.} \bibinfo{year}{2021}\natexlab{a}.
\newblock \showarticletitle{Automated Fact-Checking for Assisting Human
  Fact-Checkers}.
\newblock \bibinfo{journal}{\emph{CoRR}}  \bibinfo{volume}{abs/2103.07769}
  (\bibinfo{year}{2021}).
\newblock
\showeprint[arXiv]{2103.07769}
\urldef\tempurl%
\url{https://arxiv.org/abs/2103.07769}
\showURL{%
\tempurl}


\bibitem[Nakov et~al\mbox{.}(2021b)]%
        {DBLP:journals/corr/abs-2109-12987}
\bibfield{author}{\bibinfo{person}{Preslav Nakov}, \bibinfo{person}{Giovanni
  Da~San Martino}, \bibinfo{person}{Tamer Elsayed}, \bibinfo{person}{Alberto
  Barr{\'{o}}n{-}Cede{\~{n}}o}, \bibinfo{person}{Rub{\'{e}}n M{\'{\i}}guez},
  \bibinfo{person}{Shaden Shaar}, \bibinfo{person}{Firoj Alam},
  \bibinfo{person}{Fatima Haouari}, \bibinfo{person}{Maram Hasanain},
  \bibinfo{person}{Watheq Mansour}, \bibinfo{person}{Bayan Hamdan},
  \bibinfo{person}{Zien~Sheikh Ali}, \bibinfo{person}{Nikolay Babulkov},
  \bibinfo{person}{Alex Nikolov}, \bibinfo{person}{Gautam~Kishore Shahi},
  \bibinfo{person}{Julia~Maria Stru{\ss}}, \bibinfo{person}{Thomas Mandl},
  \bibinfo{person}{M{\"{u}}cahid Kutlu}, {and} \bibinfo{person}{Yavuz~Selim
  Kartal}.} \bibinfo{year}{2021}\natexlab{b}.
\newblock \showarticletitle{Overview of the {CLEF-2021} CheckThat! Lab on
  Detecting Check-Worthy Claims, Previously Fact-Checked Claims, and Fake
  News}.
\newblock \bibinfo{journal}{\emph{CoRR}}  \bibinfo{volume}{abs/2109.12987}
  (\bibinfo{year}{2021}).
\newblock
\showeprint[arXiv]{2109.12987}
\urldef\tempurl%
\url{https://arxiv.org/abs/2109.12987}
\showURL{%
\tempurl}


\bibitem[Nogara et~al\mbox{.}(2022)]%
        {nogara2022disinformation}
\bibfield{author}{\bibinfo{person}{Gianluca Nogara},
  \bibinfo{person}{Padinjaredath~Suresh Vishnuprasad}, \bibinfo{person}{Felipe
  Cardoso}, \bibinfo{person}{Omran Ayoub}, \bibinfo{person}{Silvia Giordano},
  {and} \bibinfo{person}{Luca Luceri}.} \bibinfo{year}{2022}\natexlab{}.
\newblock \showarticletitle{The disinformation dozen: An exploratory analysis
  of covid-19 disinformation proliferation on twitter}. In
  \bibinfo{booktitle}{\emph{14th ACM Web Science Conference 2022}}.
  \bibinfo{pages}{348--358}.
\newblock


\bibitem[Park et~al\mbox{.}(2022)]%
        {park2022voynaslov}
\bibfield{author}{\bibinfo{person}{Chan~Young Park}, \bibinfo{person}{Julia
  Mendelsohn}, \bibinfo{person}{Anjalie Field}, {and} \bibinfo{person}{Yulia
  Tsvetkov}.} \bibinfo{year}{2022}\natexlab{}.
\newblock \showarticletitle{Challenges and Opportunities in Information
  Manipulation Detection: An Examination of Wartime Russian Media}.
\newblock \bibinfo{journal}{\emph{Proceedings of the 2022 Conference on
  Empirical Methods in Natural Language Processing (EMNLP)}}
  (\bibinfo{year}{2022}).
\newblock


\bibitem[Patel(2019)]%
        {DBLP:journals/corr/abs-1908-02451}
\bibfield{author}{\bibinfo{person}{Manish Patel}.}
  \bibinfo{year}{2019}\natexlab{}.
\newblock \showarticletitle{TinySearch - Semantics based Search Engine using
  Bert Embeddings}.
\newblock \bibinfo{journal}{\emph{CoRR}}  \bibinfo{volume}{abs/1908.02451}
  (\bibinfo{year}{2019}).
\newblock
\showeprint[arXiv]{1908.02451}
\urldef\tempurl%
\url{http://arxiv.org/abs/1908.02451}
\showURL{%
\tempurl}


\bibitem[Pierri and Ceri(2019)]%
        {pierri2019false}
\bibfield{author}{\bibinfo{person}{Francesco Pierri} {and}
  \bibinfo{person}{Stefano Ceri}.} \bibinfo{year}{2019}\natexlab{}.
\newblock \showarticletitle{False news on social media: a data-driven survey}.
\newblock \bibinfo{journal}{\emph{ACM Sigmod Record}} \bibinfo{volume}{48},
  \bibinfo{number}{2} (\bibinfo{year}{2019}), \bibinfo{pages}{18--27}.
\newblock


\bibitem[Pierri et~al\mbox{.}(2023a)]%
        {pierri2023one}
\bibfield{author}{\bibinfo{person}{Francesco Pierri},
  \bibinfo{person}{Matthew~R DeVerna}, \bibinfo{person}{Kai-Cheng Yang},
  \bibinfo{person}{David Axelrod}, \bibinfo{person}{John Bryden}, {and}
  \bibinfo{person}{Filippo Menczer}.} \bibinfo{year}{2023}\natexlab{a}.
\newblock \showarticletitle{One Year of COVID-19 Vaccine Misinformation on
  Twitter: Longitudinal Study}.
\newblock \bibinfo{journal}{\emph{Journal of Medical Internet Research}}
  \bibinfo{volume}{25} (\bibinfo{year}{2023}), \bibinfo{pages}{e42227}.
\newblock


\bibitem[Pierri et~al\mbox{.}(2022)]%
        {pierri2022does}
\bibfield{author}{\bibinfo{person}{Francesco Pierri}, \bibinfo{person}{Luca
  Luceri}, {and} \bibinfo{person}{Emilio Ferrara}.}
  \bibinfo{year}{2022}\natexlab{}.
\newblock \showarticletitle{How does Twitter account moderation work? Dynamics
  of account creation and suspension during major geopolitical events}.
\newblock \bibinfo{journal}{\emph{arXiv preprint arXiv:2209.07614}}
  (\bibinfo{year}{2022}).
\newblock


\bibitem[Pierri et~al\mbox{.}(2023b)]%
        {pierri2022propaganda}
\bibfield{author}{\bibinfo{person}{Francesco Pierri}, \bibinfo{person}{Luca
  Luceri}, \bibinfo{person}{Nikhil Jindal}, {and} \bibinfo{person}{Emilio
  Ferrara}.} \bibinfo{year}{2023}\natexlab{b}.
\newblock \showarticletitle{Propaganda and Misinformation on Facebook and
  Twitter during the Russian Invasion of Ukraine}. In
  \bibinfo{booktitle}{\emph{WebSci’23 -- 15th ACM Web Science Conference}}.
\newblock


\bibitem[Reimers and Gurevych(2019)]%
        {DBLP:journals/corr/abs-1908-10084}
\bibfield{author}{\bibinfo{person}{Nils Reimers} {and} \bibinfo{person}{Iryna
  Gurevych}.} \bibinfo{year}{2019}\natexlab{}.
\newblock \showarticletitle{Sentence-BERT: Sentence Embeddings using Siamese
  BERT-Networks}. In \bibinfo{booktitle}{\emph{Proceedings of the 2019
  Conference on Empirical Methods in Natural Language Processing}}.
  \bibinfo{publisher}{Association for Computational Linguistics}.
\newblock
\urldef\tempurl%
\url{https://arxiv.org/abs/1908.10084}
\showURL{%
\tempurl}


\bibitem[Robertson and Zaragoza(2009)]%
        {10.1561/1500000019}
\bibfield{author}{\bibinfo{person}{Stephen Robertson} {and}
  \bibinfo{person}{Hugo Zaragoza}.} \bibinfo{year}{2009}\natexlab{}.
\newblock \showarticletitle{The Probabilistic Relevance Framework: BM25 and
  Beyond}.
\newblock \bibinfo{journal}{\emph{Found. Trends Inf. Retr.}}
  \bibinfo{volume}{3}, \bibinfo{number}{4} (\bibinfo{date}{April}
  \bibinfo{year}{2009}), \bibinfo{pages}{333–389}.
\newblock
\showISSN{1554-0669}
\urldef\tempurl%
\url{https://doi.org/10.1561/1500000019}
\showDOI{\tempurl}


\bibitem[Saeed et~al\mbox{.}(2022)]%
        {10.1145/3511808.3557279}
\bibfield{author}{\bibinfo{person}{Mohammed Saeed}, \bibinfo{person}{Nicolas
  Traub}, \bibinfo{person}{Maelle Nicolas}, \bibinfo{person}{Gianluca
  Demartini}, {and} \bibinfo{person}{Paolo Papotti}.}
  \bibinfo{year}{2022}\natexlab{}.
\newblock \showarticletitle{Crowdsourced Fact-Checking at Twitter: How Does the
  Crowd Compare With Experts?}. In \bibinfo{booktitle}{\emph{Proceedings of the
  31st ACM International Conference on Information \&amp; Knowledge
  Management}} (Atlanta, GA, USA) \emph{(\bibinfo{series}{CIKM '22})}.
  \bibinfo{publisher}{Association for Computing Machinery},
  \bibinfo{address}{New York, NY, USA}, \bibinfo{pages}{1736–1746}.
\newblock
\showISBNx{9781450392365}
\urldef\tempurl%
\url{https://doi.org/10.1145/3511808.3557279}
\showDOI{\tempurl}


\bibitem[Samarinas et~al\mbox{.}(2021)]%
        {samarinas-etal-2021-improving}
\bibfield{author}{\bibinfo{person}{Chris Samarinas}, \bibinfo{person}{Wynne
  Hsu}, {and} \bibinfo{person}{Mong~Li Lee}.} \bibinfo{year}{2021}\natexlab{}.
\newblock \showarticletitle{Improving Evidence Retrieval for Automated
  Explainable Fact-Checking}. In \bibinfo{booktitle}{\emph{Proceedings of the
  2021 Conference of the North American Chapter of the Association for
  Computational Linguistics: Human Language Technologies: Demonstrations}}.
  \bibinfo{publisher}{Association for Computational Linguistics},
  \bibinfo{address}{Online}, \bibinfo{pages}{84--91}.
\newblock
\urldef\tempurl%
\url{https://doi.org/10.18653/v1/2021.naacl-demos.10}
\showDOI{\tempurl}


\bibitem[Shaar et~al\mbox{.}(2022)]%
        {DBLP:journals/corr/abs-2104-07423}
\bibfield{author}{\bibinfo{person}{Shaden Shaar}, \bibinfo{person}{Firoj Alam},
  \bibinfo{person}{Giovanni Da~San~Martino}, {and} \bibinfo{person}{Preslav
  Nakov}.} \bibinfo{year}{2022}\natexlab{}.
\newblock \showarticletitle{The Role of Context in Detecting Previously
  Fact-Checked Claims}. In \bibinfo{booktitle}{\emph{Findings of the
  Association for Computational Linguistics: NAACL-HLT 2022}}
  \emph{(\bibinfo{series}{NAACL-HLT~'22})}. \bibinfo{address}{Seattle,
  Washington, USA}.
\newblock


\bibitem[Shaar et~al\mbox{.}(2020)]%
        {shaar-etal-2020-known}
\bibfield{author}{\bibinfo{person}{Shaden Shaar}, \bibinfo{person}{Nikolay
  Babulkov}, \bibinfo{person}{Giovanni Da~San~Martino}, {and}
  \bibinfo{person}{Preslav Nakov}.} \bibinfo{year}{2020}\natexlab{}.
\newblock \showarticletitle{That is a Known Lie: Detecting Previously
  Fact-Checked Claims}. In \bibinfo{booktitle}{\emph{Proceedings of the 58th
  Annual Meeting of the Association for Computational Linguistics}}.
  \bibinfo{publisher}{Association for Computational Linguistics},
  \bibinfo{address}{Online}, \bibinfo{pages}{3607--3618}.
\newblock
\urldef\tempurl%
\url{https://doi.org/10.18653/v1/2020.acl-main.332}
\showDOI{\tempurl}


\bibitem[Vo and Lee(2020)]%
        {vo2020facts}
\bibfield{author}{\bibinfo{person}{Nguyen Vo} {and} \bibinfo{person}{Kyumin
  Lee}.} \bibinfo{year}{2020}\natexlab{}.
\newblock \showarticletitle{Where Are the Facts? Searching for Fact-checked
  Information to Alleviate the Spread of Fake News}. In
  \bibinfo{booktitle}{\emph{Proceedings of the 2020 Conference on Empirical
  Methods in Natural Language Processing (EMNLP)}}.
  \bibinfo{publisher}{Association for Computational Linguistics},
  \bibinfo{address}{Online}, \bibinfo{pages}{7717--7731}.
\newblock
\urldef\tempurl%
\url{https://doi.org/10.18653/v1/2020.emnlp-main.621}
\showDOI{\tempurl}


\bibitem[Xiong et~al\mbox{.}(2020)]%
        {DBLP:journals/corr/abs-2007-00808}
\bibfield{author}{\bibinfo{person}{Lee Xiong}, \bibinfo{person}{Chenyan Xiong},
  \bibinfo{person}{Ye Li}, \bibinfo{person}{Kwok{-}Fung Tang},
  \bibinfo{person}{Jialin Liu}, \bibinfo{person}{Paul~N. Bennett},
  \bibinfo{person}{Junaid Ahmed}, {and} \bibinfo{person}{Arnold Overwijk}.}
  \bibinfo{year}{2020}\natexlab{}.
\newblock \showarticletitle{Approximate Nearest Neighbor Negative Contrastive
  Learning for Dense Text Retrieval}.
\newblock \bibinfo{journal}{\emph{CoRR}}  \bibinfo{volume}{abs/2007.00808}
  (\bibinfo{year}{2020}).
\newblock
\showeprint[arXiv]{2007.00808}
\urldef\tempurl%
\url{https://arxiv.org/abs/2007.00808}
\showURL{%
\tempurl}


\end{thebibliography}

\end{document}